\documentclass[12pt,reqno]{amsart}
\usepackage{amsthm,amsfonts,amssymb,euscript}

\newcommand{\bea}{\begin{eqnarray}}
\newcommand{\eea}{\end{eqnarray}}
\def\beaa{\begin{eqnarray*}}
\def\eeaa{\end{eqnarray*}}
\def\ba{\begin{array}}
\def\ea{\end{array}}
\def\be#1{\begin{equation} \label{#1}}
\def \eeq{\end{equation}}

\def\a{{\alpha}}

\def\b{{\beta}}
\def\be{{\beta}}
\def\ga{\gamma}
\def\Ga{\Gamma}
\def\de{\delta}

\def\ep{\epsilon}
\def\eps{\epsilon}

\def\la{\lambda}

\def\si{\sigma}

\def\Om{\Omega}

\def\ze{\zeta}

\def\nab{\partial}

\def\pr{{\partial}}
\def\al{\alpha}

\def\c{\cdot}

\def\NN{{\mathcal N}}
\def\II{{I}}

\def\FF{{\mathcal F}}

\def\SS{{\mathcal S}}
\def\NN{{\mathcal N}}

\def\KK{{\mathcal K}}
\def\Lie{{\mathcal L}}

\def\Lie{{\mathcal L}}

\def\D{{\bf D}}

\def\M{{\bf M}}

\def\O{{\bf O}}

\def\Z{{\bf Z}}
\def\R{{\bf R}}

\def\K{{\bf K}}
\def\T{{\bf T}}

\def\g{{\bf g}}

\def\f12{{\frac 1 2}}
\def\ub{\underline{u}}

\def\dual{{\,\,^*}}
\def\div{{\mbox div\,}}
\def\curl{{\mbox curl\,}}

\def\Lb{{\,\underline{L}}}

\def\NNb{\underline{\NN}}

\def\trch{{\mbox tr}\, \chi}
\def\trchb{{\mbox tr}\, \chib}

\def\chib{{\underline \chi}}

\def\bb{{\underline{\b}}}
\def\aa{{\underline{\a}}}

\def\f{\widetilde{f}}

\def\um{\underline{u}}
\def\Lp{L}
\def\Lm{\underline{L}}

\newtheorem{theorem}{Theorem}[section]
\newtheorem{lemma}[theorem]{Lemma}
\newtheorem{proposition}[theorem]{Proposition}

\newtheorem{definition}[theorem]{Definition}

\numberwithin{equation}{section}

\begin{document}\title[Hawking's Killing vector-field without analyticity]{Hawking's local rigidity theorem without analyticity}
\author{S. Alexakis}
\address{Massachusetts Institute of Technology}
\email{alexakis@math.mit.edu}
\author{A. D. Ionescu}
\address{University of Wisconsin -- Madison}
\email{ionescu@math.wisc.edu}
\author{S. Klainerman}
\address{Princeton University}
\email{seri@math.princeton.edu}

\begin{abstract}
We prove the existence of a Hawking Killing vector-field in a full neighborhood of a local, regular, bifurcate, non-expanding horizon embedded in a smooth vacuum Einstein manifold. The result extends a previous result of Friedrich,  R\'{a}cz and Wald, see \cite[Proposition B.1]{FrRaWa}, which was limited to the domain of dependence of the bifurcate  horizon.   So far,  the existence of a Killing vector-field in a full neighborhood has been proved only under the restrictive assumption of analyticity of the space-time.   We also prove that,  if  the space-time possesses  an additional  Killing vectorfield $\T$,  tangent to the horizon and not vanishing identically on the bifurcation sphere, then there  must exist a local   rotational Killing field commuting 
with $\T$. Thus the space-time must be locally axially symmetric. 
The existence of  a  Hawking vector-field  $\K$, and the  above mentioned axial symmetry,    plays a fundamental role in the classification theory of stationary  black holes.   In \cite{Al-Io-Kl}  we  use the results of this paper to prove a perturbative    version of the  uniqueness of  smooth, stationary black holes in vacuum.

\end{abstract}
\maketitle
\tableofcontents

\section{Introduction}\label{introduction}
Let  $(\M,\g)$ to be a smooth\footnote{$\M$ is assumed to be a connected, oriented, $C^\infty$ $4$-dimensional manifold without boundary.} vacuum Einstein space-time. Let $S$  be an embedded spacelike $2$-sphere in $\M$ and let $\NN, \NNb$ be the  null boundaries of the causal set of $S$, i.e. the union of the causal  future and past of $S$. We fix $\O$ to be a small neighborhood of $S$ such that both $\NN,\underline{\NN}$ are regular, achronal, null hypersurfaces in $\O$ spanned by null geodesic generators orthogonal to $S$. We say that  the triplet $(S, \NN,\underline{\NN})$ forms a local, regular, bifurcate,  non-expanding horizon  in $\O$ if both  $\NN,\underline{\NN}$ are non-expanding null hypersurfaces (see definition \ref{def:nonexp})   in $\O$. Our main result is  the following:

\begin{theorem}\label{mainfirststep}
Given a   local, regular, bifurcate, non-expanding  horizon  $(S,\, \NN,\, \underline{\NN})$ in  
 a smooth,
vacuum Einstein space-time $(\mathbf{O},\g)$,   there exists  an open neighborhood $\mathbf{O}'\subseteq\mathbf{O}$ of $S$ and a non-trivial Killing vector-field $\K$ in $\mathbf{O}'$, which is tangent to the null generators of $\NN$ and $\NNb$. In other words, every local, regular, bifurcate, non-expanding horizon is a Killing bifurcate horizon.
\end{theorem}
It is already known, see \cite{FrRaWa}, that such a Killing vector-field exists in a small neighborhood of $S$ intersected with the domain of dependence of $\NN\cup\underline{\NN}$. The extension  of $\K$ to a full neighborhood of $S$ has been known to hold only under the restrictive  additional  assumption of  analyticity of the space-time (see \cite{HaEl}, \cite{IsMon}, \cite{FrRaWa}). The novelty of our theorem is the existence of Hawking's Killing vector-field $\K$ in a full neighborhood of the 2-sphere $S$, without making any analyticity assumption. It is precisely this information, i.e. the existence of $\K$ in the complement of the domain of dependence of $\NN\cup\underline{\NN}$, that is needed in the application of Hawking's rigidity theorem to the classification theory of stationary, regular black holes. The assumption that the non-expanding horizon in Theorem \ref{mainfirststep} is {\textit{bifurcate}} is essential for the proof; this assumption is consistent with the application mentioned above.     

We also  prove the following:
\begin{theorem}
\label{thm:rotation}
 Assume that  $(S,\, \NN,\, \underline{\NN})$  is a   local, regular, bifurcate, horizon
 in  a vacuum Einstein space-time $(\mathbf{O},\g)$  which possesses a Killing 
 vectorfield $\T$ 
 tangent to  $\NN\cup\NNb$ and non-vanishing on $S$.    Then,  there exists  an open neighborhood $\mathbf{O}'\subseteq\mathbf{O}$ of $S$ and a non-trivial rotational  Killing vector-field $\Z$ in $\mathbf{O}'$ which commutes with $\T$. 
\end{theorem}
Once more,   a  related  version  of result was known only in the special  case  when the space-time is analytic. In fact S.  Hawking's famous rigidity theorem, see \cite{HaEl},  asserts  that, under some  global causality, asymptotic flatness and connectivity   assumptions, a stationary, non-degenerate   analytic  spacetime  must be  axially symmetric.   Observe that, though we have not assumed specifically that the horizon is non-expanding,  this is in fact a well known consequence
of the fact that the Killing field  $\T$ is  tangent to it. Thus, in view of Theorem \ref{mainfirststep}, 
there exists a Hawking  vectorfield $\K$, in a full neighborhood of $S$. We show that there exist   constants $\la_0$ and $t_0>0$ such that
  \bea
\Z=\T+\la_0\K
\eea
 is a rotation with period $t_0$.  The main constants $\lambda_0$ and $t_0$ can be determined on the bifurcation sphere $S$.   We remark that, though  Hawking's rigidity  theorem does not require, explicitely,   a regular, bifurcate horizon,   our assumption    is  related  to   that of the  non-degeneracy of the event horizon, see \cite{Ra-Wa}.

\medskip 

 As known the existence of the Hawking vector-field  
  plays a fundamental role in the classification theory of stationary black holes (see \cite{HaEl} or \cite{ChCo} and references therein for a more complete treatment of the problem).
 The results  of this paper are used in \cite{Al-Io-Kl} to prove   a  perturbative version, without analyticity,  of the  uniqueness of  smooth, stationary black holes in vacuum.
    More precisely we show that   a regular, smooth, asymptotically flat  solution
of the vacuum Einstein equations which is  a  perturbation of a Kerr solution  $\KK(a,m)$ with $0\le a <m$ is in fact a Kerr solution.  The perturbation condition is expressed geometrically by assuming that the Mars-Simon tensor $\SS$ of the stationary space-time (see \cite{Ma1} and  \cite{IoKl} )
is sufficiently small.  The proof uses  Theorem \ref{mainfirststep} as a first step;   one first defines a Hawking vector-field $\K$   in  a neighborhood  of $S$ and then  extends
it  to the entire space-time  by using the level sets of a  canonically defined function $y$. One can show that these level sets are conditionally pseudo-convex, as  in \cite{IoKl},  as long as the the Mars-Simon tensor $\SS$ is sufficiently small.  Once 
$\K$ is extended to the entire space-time  one can show, using the  result
 of Theorem \ref{thm:rotation},  that the space-time is  not only stationary but also axisymmetric.  The proof then follows by appealing to  the methods of the  well
known results of Carter \cite{Ca1} and Robinson \cite{Rob}, see also the more complete account \cite{ChCo}.

\medskip

\subsection{Main Ideas} 
We recall that a  Killing vector-field $\K$ in a vacuum Einstein space-time must verify the covariant
wave equation
\bea
\square_\g \K=0. \label{eq:intr1}
\eea
The main  idea in  \cite{FrRaWa} was to construct  $\K$ as a  solution to 
\eqref{eq:intr1} with appropriate, characteristic,  boundary conditions on 
$\NN\cup\NNb$.  As known, the characteristic initial value problem is well posed in the domain of dependence of $\NN\cup\NNb$ but ill posed in its complement. 
To avoid this fundamental   difficulty  we  rely instead on a completely different strategy\footnote{\ Such a strategy was discussed in \cite[Remark  B.1.]{FrRaWa}, as an alternative to  the use of the wave equation  \eqref{eq:intr1},   in the  domain 
of dependence of  $\NN\cup\NNb$. We would like to thank  R. Wald for drawing our attention to it.}.  The main idea, which allows us to avoid using \eqref{eq:intr1} or some other  system of PDE's in the ill posed region, is to first construct $\K$ in the domain of dependence of $\NN\cup\NNb$ as a solution to \eqref{eq:intr1}, extend $\K$  by Lie  dragging along the   null geodesics transversal to $\NN$,   consider its   associated flow  $\Psi_t$,   and  
 show  that, for small $|t|$, the pull back metric  $\Psi_t^*\g$ must coincide with $\g$, in view of  the fact they they are both solutions of the Einstein vacuum   equations  and   coincide on $\NN\cup\NNb$.  To implement this idea we need to prove a uniqueness result for two  Einstein vacuum metrics $\g$, $\g'$  which coincide on $\NN\cup\NNb$.      Such
 a  uniqueness result was proved by one of the authors in \cite{Al},  based on the uniqueness results for systems of covariant  wave equations proved by the other two authors in \cite{IoKl} and \cite{IoKl2}.
 The starting point of 
the proof are  the  schematic identities,
\beaa
\square_\g \R=\R* \R,\qquad \square_{\g'} \R'=\R'* \R'
\eeaa
with $\R*\R$, $\R'*\R'$   quadratic expressions in the curvatures $\R, \R'$
of  the Einstein vacuum metrics $\g, \g'$.  Subtracting the two equations we derive,
\beaa
\square_\g(\R-\R')+\big(\square_\g-\square_{\g'} \big) \R'=
(\R-\R')*(\R+\R').
\eeaa
We would like to rely on the uniqueness properties 
of covariant wave equations, as in \cite{IoKl}, \cite{IoKl2},
 but this is not possible due to 
 the presence of the term   $\big(\square_\g-\square_{\g'} \big) \R'$ which forces
 us to consider equations for  $\g-\g'$ expressed relative to an appropriate choice   of  a gauge condition.  
  An obvious such  gauge  choice  would be the wave gauge 
 $\square_\g x^\a=0$ which would lead to a system of wave 
 equations for the components of the two metrics $\g$, $\g'$
 in the given coordinate system.
 Unfortunately such coordinate system would have to be constructed
 starting with data on $\NN\cup\underline{\NN}$ which requires one to solve  the same ill posed problem.
 We  rely  instead on  a pair of   geometrically constructed     frames $v$,  $v'$ (using parallel transport with respect to $\g$ and $\g'$)  and derive ODE's for their difference  $dv=v'-v$,  as well as the difference  $d\Ga=\Ga'-\Ga$
 between their connection coefficients,  with source terms  in $d R=\R'-\R$.
 In this way we derive a  system of wave equations in $d R$
 coupled with ODE's in $dv$, $d\Ga$ and their partial  derivatives
 $\partial dv$,  $\partial d\Ga$ with respect to our fixed coordinate system. Since   ODE's are clearly well posed it is natural to expect
 that the uniqueness results for covariant wave equations derived
 in  \cite{IoKl}, \cite{IoKl2}  can be extended to such coupled system  and thus deduce that $d v=d\Ga=d R=0$ in a full neighborhood of $S$. The  precise result  is stated and proved in Lemma \ref{extendedCarl}.
 
 \medskip

 In section 2 we  construct a  canonical    null frame which will be used  throughout 
 the paper.  We  use the non-expanding condition to  derive the main null structure 
 equations  along $\NN$ and $\NNb$.  In section 3 we  give a self contained 
 proof of Proposition B.1. in \cite{FrRaWa} concerning the existence of a 
 Hawking vector-field in  the domain  of dependence of $\mathcal{N}\cup\underline{\mathcal{N}}$. In section 4,
 we show how to extend $\K$ to a full neighborhood of $S$.  We also show that the extension must be locally   time-like in the complement of   the domain of dependence of $\mathcal{N}\cup\underline{\mathcal{N}}$, see Proposition \ref{Ktimelike}.  In section 5 we  prove Theorem 
 \ref{thm:rotation}. We  first   show that if  $\T$ is another smooth Killing vector-field, tangent
 to $\NN\cup \NNb$, then it must commute with $\K$ in a full neighborhood of $S$. We then construct a rotational Killing vector-field $\Z$   as a linear combination of $\T$ and $\K$.  We also
 show that if  $\si_\mu$ is the Ernst potential associated with $\T$ then  $\K^\mu=\Z^\mu \si_\mu=0$. 
 These additional results, in the presence of the (stationary) Killing vector-field $\T$, are important in the application in \cite{Al-Io-Kl}. 
 
 \medskip

{\bf Acknowledgements}:\quad We would like to thank   P. Chrusciel, M. Dafermos and   R. Wald for  helpful discussions and suggestions.

\section{Preliminaries}\label{prelim}
We restrict our attention to an open neighborhood $\O$
of $S$ in which   $\NN,\underline{\NN}$   are regular, achronal,  null hypersurfaces, spanned by null geodesic generators orthogonal to     $S$.   During the proof
of our main theorem and their consequences we will keep restricting  our attention to smaller and smaller neighborhoods of $S$; for simplicity of notation we keep denoting such neighborhoods of $S$ by $\O$. 

We define two optical functions $u,\underline{u}$ in a neighborhood of $S$ as follows. We first  fix a smooth future-directed  null pair $(L, \Lb)$ along $S$, satisfying
\begin{equation}\label{normalization}
\g(L,L)=\g(\Lb,\Lb)=0,\,\,\,\,\,\g(L, \Lb)=-1,
\end{equation}
such that $L$ is tangent to $\NN$ and $\Lb$ is tangent to $\underline{\NN}$. In a small neighborhood of $S$, we extend $L$ (resp. $\Lb$)  along the null geodesic generators of $\NN$
 (resp. $\underline{\NN}$) by parallel transport, i.e. $\D_LL=0$ (resp.  $\D_\Lb\Lb=0$).  
 We define the function $\underline{u}$ (resp. $u$) along $\NN$ (resp. $\underline{\NN}$) by setting $u=\underline{u}=0$ on $S$ and solving $L(\underline{u})=1$ (resp.  $\Lb(u)=1$).  Let $S_{\underline{u}}$ (resp.  $\underline{S}_{u}$)  be the level surfaces of $\underline{u}$ (resp. $u$)   along $\NN$ (resp. $\underline{\NN}$). We define $\Lb$ at every point of $\NN$ (resp. $L$ at every point of $\underline{\NN}$) as the unique, future directed null vector-field orthogonal to the surface $S_{\underline{u}}$ (resp. $\underline{S}_u$) passing through that point and such that $\g(L,\Lb)=-1$.
 We now define the null hypersurface $\underline{\NN}_{\underline{u}}$ to be the congruence
 of   null geodesics  initiating on $S_{\underline{u}}\subset\NN$ in the direction of $\Lb$.
 Similarly we define $\NN_u$ to be the congruence
 of   null geodesics  initiating on $\underline{S}_{u}\subset\underline{\NN}$ in the direction of  $L$.
 Both congruences are well defined in a sufficiently small neighborhood of $S$ in $\O$, which (according to our convention) we continue to call $\O$.  The null hypersurfaces $\underline{\NN}_{\underline{u}}$ (resp.  $\NN_{u}$) are the level sets
 of  a function $\underline{u}$  (resp $u$)  vanishing on $\underline{\NN}$  (resp. $\NN$). By construction 
  \begin{equation}\label{haw8}
  L=-\g^{\mu\nu}\pr_\mu u\pr_\nu,\qquad  \Lb=-\g^{\mu\nu}\pr_\mu\underline{u}\pr_\nu.
  \end{equation}
In particular, the functions $u,\underline{u}$ are both null optical functions,
 i.e.
 \begin{equation}\label{haw9}
 \g^{\mu\nu}\pr_\mu u \pr_\nu u=\g(L,L)=0\quad\text{ and }\quad \g^{\mu\nu}\pr_\mu \underline{u}\, \pr_\nu \underline{u}=\g(\Lb,\Lb)=0.
 \end{equation}
 We  define,
 \begin{equation*}
 \Om=\g^{\mu\nu}\pr_\mu u\,  \pr_\nu \underline{u}=\g(L,\Lb).
 \end{equation*}  
By construction $\Omega=-1$ on $(\NN\cup\underline{\NN})\cap\O$, but $\Omega$ is not necessarily equal to $-1$ in $\O$. Choosing 
 $\O$  small enough, we may assume however  that $\Omega\in[-3/2,-1/2]$ in $ \O$.

To  summarize,  we can find  two 
smooth optical functions $u,\underline{u}:\O\to\mathbb{R}$ such that, 

\begin{equation}\label{haw11}
\NN\cap\O=\{p\in\O:u(p)=0\},\qquad\underline{\NN}\cap\O=\{p\in\O:\underline{u}(p)=0\}.
\end{equation}
and,
\begin{equation}\label{haw12}
\Omega\in[-3/2,-1/2]\quad\text{ in }\O.
\end{equation}
Moreover,  by construction (with  $L,\Lb$ defined by  \eqref{haw8})  we have,  
\beaa
L(\underline{u})=1\,\, \mbox{on}\,\,\,  \NN,\qquad \Lb(u)=1 \,\,\, \mbox{on} \,\,
\underline{\NN}.
\eeaa
Using the null pair $\Lb, L$  introduced in \eqref{normalization}, \eqref{haw8} we fix an associated  null frame $e_1, e_2, e_3=\Lb, e_4=L$ such that $\g(e_a,e_a)=1$, $\g(e_1,e_2)=\g(e_4,e_a)=\g(e_3,e_a)=0$, $a=1,2$.  At every point $p$  in  in $\O$,    $e_1$, $e_2$ form an orthonormal
frame along the $2$-surface $S_{u, \underline{u} }$  passing through $p$. We denote 
by $\nabla$ the induced covariant derivative operator on $S_{u,\underline{u}}$.  Given a horizontal vector-field  $X$, i.e. $X$ tangent to the $2$-surfaces  $S_{u,\underline{u}}$ at every point in  $\O$, we  
denote by $\nabla_3 X$, $\nabla_4X $ the projections  of $\D_{e_3}$ and $\D_{e_4}$
to   $S_{u,\underline{u}}$. 
 Recall the definition of the null second fundamental forms
\begin{equation*}
\chi_{ab}=\g(\D_{e_a}L, e_b),\qquad \chib_{ab}=\g(\D_{e_a}\Lb, e_b)
\end{equation*}
and the torsion
\begin{equation*}
\ze_a=\g(\D_{e_a}L, \Lb).
\end{equation*}
\begin{definition}
\label{def:nonexp}
We say that $\NN$   is    non-expanding if $\trch=0$ on $\NN$.  Similarly  $\NNb$
is non-expanding if  $\trchb=0$ on $\NNb$. The bifurcate horizon $(S, \NN, \NNb)$
is called non-expanding if both $\NN, \NNb$ are non-expanding. 
\end{definition}

The assumption that the surfaces $\NN$ and $\underline{\NN}$ are non-expanding   implies,
according to the Raychadhouri equation, 
\begin{equation}\label{Haw15}
\chi=0\text{ on }\NN\cap\O\quad\text{ and }\quad \chib=0\text{ on }\underline{\NN}\cap\O.
\end{equation}
In addition, since the vectors $e_1,e_2$ are tangent to $(\NN\cup\underline{\NN})\cap\O$ and $\g(L,\Lb)=-1$ on $(\NN\cup\underline{\NN})\cap\O$, we have $\ze_a=-\g(D_{e_a}\Lb,L)$ on $(\NN\cup\underline{\NN})\cap\O$. Finally, it is known that  the following components of the curvature tensor $\R$ vanish on $\NN$ and $\underline{\NN}$, 
\begin{equation}\label{Haw16}
\R_{4a4b}=\R_{434b}=0\text{ on }\NN\quad\text{ and }\quad \R_{3a3b}=\R_{343b}=0\text{ on }\underline{\NN},\quad a,b=1,2.
\end{equation}
Let, see  \cite{CKl}, \cite{KlNi},   $\a_{ab}=\R_{4a4b}$, $\b_a= \R_{a434}$,
$\rho=\R_{3434}$, $\si=\dual \R_{3434}$, $\bb_a=\R_{a334}$ and $\aa_{ab}=\R_{a3b3}$ denote  the null components of $\R$. Thus,  in view of \eqref{Haw16} the only non-vanishing null  components of $\R$ on $S$ are $\rho$ and $\si$.
 Since $[e_a,e_4](\underline{u})=0$ on $\NN\cap\O$, it follows that $\g([e_a,e_4],e_3)=0$ on $\NN\cap\O$. Using $\D_LL=0$, \eqref{Haw15}, and the definitions, we derive, on $\NN\cap\O$,
\begin{equation}\label{table1}
\begin{split}
&\D_{e_4}e_4=0,\quad\D_{e_a}e_4=-\ze_a e_4,\quad\D_{e_4}e_3=-\sum_{a=1}^2\ze_b e_b,\quad \D_{e_4}e_a=\nabla_{e_4}e_a-\ze_a e_4,\\
&\D_{e_a}e_3=\sum_{b=1}^2\chib_{ab}e_b+\ze_ae_3,\quad \D_{e_a}e_b=\nabla_{e_a}e_b+\chib_{ab}e_4.
\end{split} 
\end{equation}

\begin{lemma}\label{nstr}
The null structure equations  along  $\NN$ (see\footnote{The discrepancy  with the corresponding formula is due to the different normalization for $\Lb$, i.e.
 $\g(L,\Lb)=-1$ instead of  $\g(L,\Lb)=-2$ . }
 Proposition 3.1.3 in  \cite{KlNi})  reduce to
\begin{equation}\label{table:null-str}
\begin{split}
&\nabla_4 \ze=0,\quad \curl \ze=\si,\quad  L(\trchb)+\div \ze-|\ze|^2=\rho. \end{split} 
\end{equation}
Also, if $X$ is an horizontal vector,
\beaa
[\nabla_4,  \nabla_a]  X_b=0.
\eeaa
As a consequence  we also have,
\bea
\nabla_4(\div \ze)=0.
\eea
\end{lemma}

\begin{proof}[Proof of Lemma \ref{nstr}]
Indeed, 
\beaa
\g(\D_4\D_a \Lb, e_4)-\g(\D_a\D_4 \Lb, e_4)=\R( e_a, e_4, e_3, e_4)=\b_a
\eeaa
and, using \eqref{table1},
$
\g(\D_a \D_4\Lb, e_4)=\Lb_{4;4a}=0$,
$\g(\D_4 \D_a\Lb, e_4)=\Lb_{4;a4}=-\nabla_4\ze_a$.
Hence, since $\b$ vanishes along $\NN$, we deduce
$
\nabla_4\ze=0.
$
Also,
\beaa
\g(\D_4\D_b\Lb,  e_a)-\g(\D_b\D_4\Lb,  e_a)&=&\R(e_a, e_3, e_4, e_b)=\frac 1 2\,\ga_{ab}\, \rho-\frac 1 2\,\in_{ab}  \si
\eeaa
and,
$\g(\D_4\D_b\Lb,  e_a)=\Lb_{a;b4}=\nabla_4\chib_{ab}-2\ze_a\ze_b$, 
$g(\D_b\D_4\Lb,  e_a)=\Lb_{a;4b}=-\nabla_b\ze_a-\ze_a\ze_b$.
Hence,
$$
\nabla_4\chib_{ab}-\ze_a\ze_b+\nab_b\ze_a=\frac 12 \rho\,\ga_{ab}-\frac 1 2 \si\in_{ab}.
$$
Taking the symmetric part we derive,
$
\nabla_4\trchb-|\ze|^2+\div \ze =\rho
$
while taking the antisymmetric part  yields,
$
\curl \ze =\si
$
as desired.
To check the commutation formula we write,
\beaa
\D_4 \D_a X_b &=& e_4(\D_a X_b)-\D_{\D_4 e_a} X_b-\D_aX_{\D_4 e_b}\\
&=&e_4(\nabla_b X_a)-\D_{\nabla_4 e_a} X_b+\ze_a \D_4 X_b-\D_aX_{\nabla_4 e_a}+\ze_b\D_a X_4\\
&=&\nabla_4\nabla_a X_b +\ze_a\nabla_4 X_b \\
 \D_a \D_4  X_b &=&e_a( \D_4  X_b)- \D_{\D_ae_4}   X_b- \D_4  X_{\D_a e_b }\\
 &=&e_a( \D_4  X_b)- \D_{\nabla_ae_4}   X_b+\ze_a \D_4 X_b-\D_4  X_{\nabla_a e_b }\\
 &=&\nabla_a\nabla_4 X_b+\ze_a \nabla_4X_b
\eeaa
Therefore,
\beaa
[\D_4, \D_a] X_b=[\nabla_4, \nabla_a] X_b.
\eeaa
On the other hand,  $[\D_4, \D_a] X_b=\R_{a4cb} X^c=0$ in view of the vanishing
of $\b$ and the Einstein equations. 
\end{proof}
We define the following  four regions $\II^{++}$, $\II^{--}$, $\II^{+-}$ and $\II^{-+}$:
\begin{equation}\label{setup1}
\begin{split}
&\II^{++}=\{p\in\O:u(p)\geq 0\text{ and }\underline{u}(p)\geq 0\},\quad \II^{--}=\{p\in\O:u(p)\leq 0\text{ and }\underline{u}(p)\leq 0\},\\
&\II^{+-}=\{p\in\O:u(p)\geq 0\text{ and }\underline{u}(p)\leq 0\},\quad \II^{-+}=\{p\in\O:u(p)\leq 0\text{ and }\underline{u}(p)\geq 0\}.
\end{split}
\end{equation}
Clearly $\II^{++}, \II^{--} $ coincide with the causal and  future and past sets of $S$ in $\O$. 
\section{Construction of the Hawking vector-field  in the causal  region} \label{Known}

We construct first the Killing vector-field $\K$ in the causal  region $\II^{++}\cup \II^{--}$.
\begin{proposition}\label{Friedrich}
Under the assumptions of Theorem \ref{mainfirststep},  there is  a small  neighborhood  $\O$ of $S$,  a smooth Killing vector-field $\K$ in $\O\cap\big(\II^{++}\cup \II^{--}\big)$ such that
\begin{equation}\label{Haw5}
\K=\underline{u}L-u\Lb\quad\text{ on }(\NN\cup\underline{\NN})\cap\O.
\end{equation} 
Moreover,  in the region $\O\cap (\II^{++}\cup \II^{--})$  where $\K$ is defined, $[\Lb,\K]=-\Lb$.
\end{proposition}

The rest  of this section is concerned with the proof of Proposition \ref{Friedrich}. The first part of the  proposition, which depends on the main assumption that the surfaces $\NN$ and $\underline{\NN}$ are non-expanding, is well  known, see \cite[Proposition B.1.]{FrRaWa}. For the sake of completeness, we provide its proof below.  

 Following \cite{FrRaWa}  we construct the smooth vector-field $\K$ as the solution to the characteristic initial-value problem,
\begin{equation}\label{Haw11}
\square_\g \K=0,\qquad \K=\underline{u}L-u\Lb\quad\text{ on }(\NN\cup\underline{\NN})\cap\O.
\end{equation}
As well  known, see \cite{Re}, the characteristic initial value problem   for  wave equations of type  \eqref{eq:A1}   is well posed.
Thus  the  vector-field $\K$ is well-defined and smooth in the domain of dependence of $\NN\cup\underline{\NN}$ in $\O$. Let $\pi_{\al\be}={}^{(\K)}\pi_{\al\be}=\D_\al\K_\be+\D_\be\K_\al$. We have to prove that $\pi=0$ in a neighborhood of $S$ intersected to  $\II^{++}\cup \II^{--}$. It follows from \eqref{Haw11}, using the Bianchi identities and the Einstein vacuum equations, that $\pi$ verifies the  covariant wave equation,
\begin{equation}
\square_\g\pi_{\al\be}=2{{\R^{\mu}}_{\al\be}}^\nu\pi_{\mu\nu}.\label{eq:A1}
\end{equation}
In view of the standard uniqueness result  for characteristic initial value problems,  see \cite{Re},    the statement of the proposition reduces  to showing  that $\pi=0$ on $(\NN\cup\underline{\NN})\cap\O$.
By symmetry, it suffices to prove that $\pi=0$ on $\NN\cap\O$.  The proof relies on   our  main hypothesis,   that the surfaces $\NN$ and $\underline{\NN}$ are non-expanding.

Since $\K=\underline{u}L$ on $\NN\cap\O$ is tangent to the null generators of $\NN$, it follows that
\begin{equation}\label{Haw17}
\D_4\K_3=-1,\quad\D_4\K_4=\D_a\K_4=\D_4\K_a=\D_a\K_b=0,\quad a,b=1,2.
\end{equation}
Thus, on $\NN\cap\O$
\begin{equation}\label{Haw18}
\pi_{44}=\pi_{a4}=\pi_{ab}=0\quad a,b=1,2.
\end{equation}
To prove that the remaining components of $\pi$ vanish we use the wave equation $\square_\g\K=0$, which gives
\begin{equation*}
\D_3\D_4\K_\mu+\D_4\D_3\K_\mu=\sum_{a=1}^2\D_a\D_a\K_\mu\quad\text{ on }\NN\cap\O.
\end{equation*}
Since $\D_3\D_4\K_\mu-\D_4\D_3\K_\mu=\R_{34\mu\nu}\K^\nu$ on $\NN\cap\O$ (using \eqref{Haw16}), we derive
\begin{equation}\label{Haw19}
2\D_4\D_3\K_\mu=\sum_{a=1}^2\D_a\D_a\K_\mu-\R_{34\mu\nu}\K^\nu,\,\mu=1,2,3,4,\text{ on }\NN\cap\O.
\end{equation}
We set first $\mu=4$. It follows from \eqref{Haw17} that $\D_4\D_3\K_4=0$. In addition, $\D_3\K_4=1$ on $S$ (the analogue of the first identity in \eqref{Haw17} along the hypersurface $\underline{\NN}$). Using \eqref{table1} and \eqref{Haw17}, $\D_4\D_3\K_4=L(\D_3\K_4)$. Thus $\D_3\K_4=1$ on $\NN$, which implies
\begin{equation}\label{Haw25}
\pi_{34}=0\quad\text{ on }\NN.
\end{equation}

We use now the equation \eqref{Haw19} with $\mu=a\in\{1,2\}$ to calculate $P_a:=\pi_{a3}$ along $\NN$. It follows from \eqref{Haw17} and \eqref{Haw16} that $\D_a\D_b\K_c=0$, $a,b,c=1,2$, and $\R_{34a\nu}\K^\nu=0$ on $\NN$. A simple computation shows that $\D_a \K_3=\underline{u}\ze_a$, thus $P_a=\D_3 \K_a+\underline{u}\ze_a$. Thus, using \eqref{table1}, $\D_3\K_4=1$, and $\D_b\K_c=0$ on $\NN$, we derive
\begin{equation*}
\begin{split}
 0&=\K_{b;34}=e_4(\K_{b;3})-\K_{\D_{e_4}e_b; e_3}-\K_{e_b;\D_{e_4}e_3}=e_4( P_b-\underline{u} \ze_b)-\K_{\nabla_{4}e_b; e_3}+\ze_b K_{e_4;e_3}\\
&=\nabla_4(P_b-\um\ze_b)+\ze_b=\nabla_4 P_b-\um \nabla_4 \ze_b.
\end{split}
\end{equation*}
Thus
 \beaa
 \nabla_4 P_a=\um \nabla_4 \ze_a\quad\text{ on }\NN.
 \eeaa
 On the other hand, along $\NN$, $\ze$ verifies the transport equation,
 \beaa
 \nabla_4\ze_a=-\R_{a434}=0.
 \eeaa
 Therefore, along $\NN$,
 \beaa
 \nabla_4 P_a=0.
 \eeaa
Since $P_a=\pi_{a3}=0$ on $S$ it follows that
\begin{equation}\label{Haw30}
\pi_{a3}=0\quad\text{ on }\NN.
\end{equation}
 
 Similarly, denoting $Q=\pi_{33}=2\D_3 \K_3$, we have, according to \eqref{Haw19} with $\mu=3$,
 \bea
 \D_4 \D_3 \K_3 =\frac 1 2\big( \sum_{a=1}^2 \D_a \D_a \K_3-   \rho  \um \big),\qquad 
 \rho=\R_{3434}.
 \label{eq:waveK3}
 \eea
 Now, since we already now that $\pi_{3b}$ vanishes on $\NN$,
 \bea
 \K_{3;34}=e_4 (\K_{3;3})-\K_{\D_{e_4} e_3; e_3}-\K_{e_3;\D_{e_4}e_3}=\frac 1 2 e_4 (Q)+\sum_{b=1}^2\ze_b\pi_{3b}=\frac 1 2 e_4(Q).\label{eq:D2D3K3} 
 \eea
 On the other hand, using \eqref{table1}, $\K_{3;4}=-1$, $\K_{a;b}=0$, and $\K_{3;a}  =\underline{u} \ze_a$,
 \begin{equation*}
\begin{split}
 \K_{3;ab}&=e_b(\K_{3;a})-\K_{e_3; \D_{e_b}e_a}-\K_{\D_{e_b}e_3;e_a}\\
&=\nab_b(\um \ze_a)   -      \chib_{ba} \K_{3;4}-\ze_b \K_{3;a}\\
&=\nab_b (\um \ze_a)    +        \chib_{ba}-\um\ze_a \ze_b,
\end{split} 
 \end{equation*}
thus
 \bea
 \sum_{a=1}^2\D_a \D_a\K_3&=\underline{u}(\div \ze-|\ze|^2)+\trchb. \label{eq:DaDaK3}
 \eea
 Therefore, equation \eqref{eq:waveK3} takes the form
 \bea
 L( Q )  &=& \trchb +\um(\div \ze-|\ze|^2-\rho). \label{eq:LpQ}
 \eea
 On the other hand we have the following  structure equation on $\NN$,
 \bea
L( \trchb)+\div \ze- |\ze|^2- \rho=0\label{eq:strtrchb}.
 \eea
 Thus, differentiating \eqref{eq:LpQ} with respect to $L$  and applying
 \eqref{eq:strtrchb} we  derive,
 \beaa
  L(L(Q))  &=&L( \trchb) +(\div \ze-|\ze|^2-\rho)+\um \Lp(\div \ze-|\ze|^2-\rho)\\
  &=&-\div \ze+ |\ze|^2+ \rho +(\div \ze-|\ze|^2-\rho)+\um \Lp(\div \ze-|\ze|^2-\rho).
 \eeaa
 Using null structure equations, it is not hard to check that 
\begin{equation}\label{tojustify}
L(\div \ze)=L(|\ze|^2)=L(\rho)=0\quad\text{ along }\NN.
\end{equation}
Indeed, the last identity follows from \eqref{Haw16} and \cite[Proposition 3.2.4]{KlNi}. The identity $L(|\ze|^2)=0$ follows from the transport equation $\nabla_4\ze_a=0$.
Therefore,
 \beaa
 L(L(Q))  &=&0  \qquad \mbox{along} \quad \NN.
 \eeaa
 Since   $ L(Q)=0$ on $S$   (using    again\eqref{eq:LpQ} restricted to $S$ where both $\trchb$ and $\um$ vanish),   we infer that $L(Q)=0$ along $\NN$.  Since $Q=0$ on $S$ we conclude   that $Q=0$ along  $\NN$ as desired. Thus $\pi_{33}=0$, as desired.

The second part of the proposition,
 $[\Lb,\K]=-\Lb$ in a neighborhood of $S$ in $I^{++}\cup I^{--}$,  follows from the identity,
\beaa
\D_\Lb W=-\D_W \Lb \quad  \mbox{where}  \quad W=[\Lb,\K]+\Lb=-\Lie_\K\Lb+\Lb,
\eeaa
and the vanishing of $W$
 on $\NN\cap\O$.
 To prove the identity we make use of the fact that $\Lie_\K$
 commutes with covariant differentiation. In particular,
 if $\K$ is Killing and $X, Y$  arbitrary vector-fields then,
 \bea
 \Lie_\K(\D_X Y)&=&\D_X (\Lie_\K Y)+     \D_{\Lie_\K X}Y.\label{eq:Lie.ident}
 \eea
 Therefore,
 \beaa
 \D_\Lb W&=&\D_\Lb(-\Lie_\K \Lb+\Lb)=-\D_\Lb\Lie_\K \Lb=\Lie_\K (\D_\Lb \Lb)+\D_{(\Lie_\K\Lb)} \Lb=-\D_W\Lb.
  \eeaa
as stated.
It remains to prove that 
\begin{equation}\label{hw00}
W=[\Lb,\K]+\Lb=0\qquad\text{ on }\NN\cap\O. 
\end{equation}
Since $\K=\underline{u}L$ on $\NN\cap\O$, this is equivalent to
\begin{equation}\label{iden2}
\D_3\K_\mu-\underline{u}\D_4\Lb_\mu+\Lb_\mu=0\text{ on }\NN\cap\O,\quad\mu=1,2,3,4.
\end{equation}
We  check \eqref{iden2} on the null frame $e_1,e_2,e_3=\Lb,e_4=L$ defined earlier. The identity \eqref{iden2} follows for $\mu=a=1,2$ since $\D_3\K_a=-\D_a\K_3=-\underline{u}\ze_a$, $\D_4\Lb_a=\g(e_a,\D_{e_4}e_3)=-\ze_a$ (see \eqref{table1}), and $\Lb_a=0$. The identity also follows for $\mu=3$ since $\D_3\K_3=\pi_{33}/2=0$ (in view of Proposition \ref{Friedrich}), $\D_4\Lb_3=\g(e_3,\D_{e_4}e_3)=0$ (see \eqref{table1}), and $\Lb_3=0$. Finally, for $\mu=4$, $\D_3\K_4=-\D_4\K_3=1$ (see \eqref{Haw17}), $\D_4\Lb_4=\g(e_4,\D_{e_4}e_3)=0$, and $\Lb_4=-1$. This completes the proof of the proposition.

\section{Extension of the Hawking vector-field  to a full neighborhood}\label{Mainnew}
In the previous section we have defined our Hawking vector-field $\K$ in a neighborhood $\O$ of $S$ intersected with $\II^{++}\cup\II^{--}$.   To extend  $\K$ in the  exterior region  $\II^{+-}\cup\II^{-+}$ we cannot rely on solving
 equation \eqref{Haw11}; the characteristic initial value problem is ill posed in that region.  We need to rely instead on  a completely different strategy, sketched in the introduction.  We  extend $\K$  by Lie  dragging it  relative to $\Lb$ and show that, for small $|t|$, $\Psi_t^*\g$ must coincide with $\g$, where $\Psi_t=\Psi_{t, \K}$ is the flow generated by $\K$. We show that  both metrics coincide on $\NN\cup\NNb$ and,
  since they  both verify the vacuum Einstein equations, we prove that the must
  coincide in a full neighborhood of $S$. 
  
To implement  this strategy 
we first  define the  vector-field $K'$ by setting $K'=\underline{u}L$ on $\NN\cap\O$ and solving the ordinary differential equation $[\Lb,K']=-\Lb$. The vector-field $K'$ is well-defined and smooth in a small neighborhood of $S$ (since $\Lb\neq 0$ on $S$)  and 
coincides with $\K$ in $I^{++}\cup I^{--}$  in $\O$. Thus $\K:=K'$ defines the desired extension. This proves the following.
\begin{lemma}\label{extendK}
There  exists   a smooth extension of the vector-field $\K$ (defined in Proposition \ref{Friedrich}) to  an open neighborhood $\O$ of $S$   such that
\begin{equation}\label{constructK}
[\Lb,\K]=-\Lb\quad\text{ in }\O.
\end{equation}
\end{lemma}

It remains to prove that  $\K$ is indeed our desired  Killing vector-field. For $|t|$ sufficiently small, we define, in a small neighborhood of $S$, the map $\Psi_t=\Psi_{t,\K}$ obtained by flowing a parameter distance $t$ along the integral curves of $\K$. Let
\begin{equation*}
\g^t=\Psi_{t}^\ast(\g).
\end{equation*}
The Lorentz metrics $\g^t$ are well-defined in a small neighborhood of $S$, for $|t|$ sufficiently small. To show
that $\K$ is Killing we need to show that in fact  $\g^t=\g$.
 Since $\K$ is tangent to  $(\NN\cup\underline{\NN})\cap\O$ and  is  Killing  in $I^{++}\cup I^{--}$, we infer  that $\g^t=\g$ in a small neighborhood of $S$ intersected with $I^{++}\cup I^{--}$. 
 In view of the definition of $\K$ (see \eqref{constructK}),   
\beaa
\frac{d}{dt}\Psi_t^* \Lb&=&\lim_{h\to 0}\frac{\Psi_{t-h}^* \Lb-
\Psi_t^*\Lb}{-h}=-\Psi_t^*\big( \lim_{h\to 0}\frac{\Psi_{-h}^* \Lb-
\Psi_0^*\Lb}{h}\big)=-\Psi_t^*(\Lie_\K\Lb)=-\Psi_t^*\Lb.
\eeaa
We infer that,
\beaa
\Psi_t^* \Lb=e^{-t} \Lb.
\eeaa
Now, given arbitrary vector-fields $X, Y$, we have
$
\D^t_{X^t} Y^t=\Psi_t^*(\D_XY)
$
where  $\D^t$ denotes the covariant derivative induced by the metric $\g^t=\Psi_t^* g$ and $X^t=\Psi_t^*X$,  $Y^t=\Psi_t^*Y$.
In, particular $ 0= \D^t_{\Lb^t}\Lb^t=e^{-2t} \D^t_\Lb \Lb$. This
proves the following.
\begin{lemma}\label{le:DtLbLb}
Assume $\K$ is a smooth vector-field
verifying \eqref{constructK} and  $\D^t$  the covariant derivative induced by the metric $\g^t=\Psi_t^* g$. Then,
 \beaa
 {\D^t}_{\Lb}\Lb=0\quad\text{ in a small neighborhood of }S.
 \eeaa
 \end{lemma}
 To summarize we have a family of metrics $\g^t$  which
 verify the Einstein vacuum equations $\mbox{\bf Ric}(\g^t)=0$,  $\g^t=\g$  in a small neighborhood of  $S$ intersected
  with $I^{++}\cup I^{--}$, and such that $ {\D^t}_{\Lb}\Lb=0$.    Without loss of generality we may assume that both relations hold in $\O$. 
  Thus Theorem \ref{mainfirststep} is an immediate  consequence of the following:

\begin{proposition}\label{mainnew}
Assume $\g'$ is a smooth Lorentz metric on $\O$,  such that $(\O, \g')$ is a smooth Einstein vacuum space-time. Assume that
\begin{equation*}
\g'=\g\text{\,\,\,  in\,\, }(\II^{++}\cup I^{--})\cap\O \quad\text{ and }\quad\D'_{\Lb}\Lb=0\text{ in }\O,
\end{equation*}
where $\D'$ denotes the covariant derivative induced by the metric $\g'$. Then $\g'=\g$ in  a small neighborhood   $\O'\subset\O $  of $S$. 
\end{proposition}
  As explained in the introduction,  this proposition  was  first proved  in \cite{Al}. We provide here a  more direct, simpler proof, specialized to our situation and based on the  uniqueness result  in Lemma \ref{extendedCarl} below. That lemma  is an extension of the uniqueness results proved in \cite{IoKl} to coupled systems of 
covariant wave equations and ODE's.  The motivation for  the proof below was given in the introduction.
\begin{proof}[Proof of Proposition \ref{mainnew}]
It suffices to prove the proposition in   a neighborhood  $\O(x_0)$  of a point $x_0$ in $S$
in which we can introduce a  fixed   coordinate system $x^\a$. Without loss of generality we may assume that 
\begin{equation}\label{haw1}
\g_{ij}(x_0)=\mathrm{diag}(-1,1,1,1),\qquad \sup_{x\in\O(x_0)}\sum_{j=0}^6 |\partial^j\g(x)|\leq A,
\end{equation}
with $|\partial^j\g|$ denoting the sum of the absolute values of all partial derivatives of order $j$ for all components of $\g$ in the given coordinate system. We may also assume, for the optical functions $u,\ub$ introduced in section 2, \begin{equation}\label{haw10}
\sup_{x\in\O(x_0)}(|\partial^ju(x)|+|\partial^j\underline{u}(x)|)\leq C_1=C_1(A)\quad\text{ for }j=0,\ldots,4.
\end{equation}
In the rest of the proof we will keep restricting to smaller and smaller neighborhoods of $x_0$; for simplicity of notation we keep denoting such neighborhoods by $\O(x_0)$.

Consider now  our old null frame $\widetilde{v}_{(1)}=e_1,\widetilde{v}_{(2)}=e_2,\widetilde{v}_{(3)}=L,\widetilde{v}_{(4)}=\Lb$ on $\NN\cap\O(x_0)$ and define the vector-fields $v_{(1)}$, $v_{(2)}$ ,$v_{(3)}$, $v_{(4)}=\Lb$ and ${v'}_{(1)}$, ${v'}_{(2)}$, ${v'}_{(3)}$, ${v'}_{(4)}=\Lb$   by parallel transport  along $\Lb$:
\begin{equation*}
\begin{split}
&\D_{\Lb}v_{(a)}=0     \text{ and }v_{(a)}=\widetilde{v}_a   \text{ on }\NN\cap\O(x_0);\\
&\D'_{\Lb}{v'}_{(a)}=0   \text{ and }{v'}_{(a)}=\widetilde{v}_a\text{ on }\NN\cap\O(x_0).
\end{split}
\end{equation*}
The vector-fields $v_{(a)}$ and ${v'}_{(a)}$ are well-defined and smooth in $\O(x_0)$. Let $\g_{(a)(b)}=\g(v_{(a)},v_{(b)})$, $\g'_{(a)(b)}=\g'({v'}_{(a)},{v'}_{(b)})$. The identities $\D_{\Lb}v_{(a)}=\D'_{\Lb}{v'}_{(a)}=0$ show that $\Lb(\g_{(a)(b)})=\Lb(\g'_{(a)(b)})=0$. Since $\g_{(a)(b)}=\g'_{(a)(b)}$ along $\NN$ it follows  that
\begin{equation}\label{a9}
\g_{(a)(b)}=\g'_{(a)(b)}:=h_{(a)(b)}\,\text{ and }\,\Lb(h_{(a)(b)})=0\text{ in }\O(x_0).
\end{equation}
For $a,b,c=1,\ldots 4$ let
\begin{equation*}
\begin{split}
&\Gamma_{(a)(b)(c)}=\g(v_{(a)},\D_{v_{(c)}}v_{(b)}),\qquad\Gamma'_{(a)(b)(c)}=\g'({v'}_{(a)},\D'_{{v'}_{(c)}}{v'}_{(b)}),\\
&(d\Gamma)_{(a)(b)(c)}=\Gamma'_{(a)(b)(c)}-{\Gamma}_{(a)(b)(c)}.
\end{split}
\end{equation*}
For $a,b,c,d=1,\ldots,4$ let
\begin{equation*}
\begin{split}
&\R_{(a)(b)(c)(d)}=\R(v_{(a)},v_{(b)},v_{(c)},v_{(d)}),\qquad{\R'}_{(a)(b)(c)(d)}=\R'({v'}_{(a)},{v'}_{(b)},{v'}_{(c)},{v'}_{(d)}),\\
&(dR)_{(a)(b)(c)(d)}={\R'}_{(a)(b)(c)(d)}-\R_{(a)(b)(c)(d)}.
\end{split}
\end{equation*}
Clearly, $\Gamma_{(a)(b)(4)}=\Gamma'_{(a)(b)(4)}=0$. We use now the definition of the Riemann curvature tensor to find a system of equations for $\Lb[(d\Gamma)_{(a)(b)(c)}]$. We have
\begin{equation*}
\begin{split}
\R_{(a)(b)(c)(d)}&=\g(v_{(a)},\D_{v_{(c)}}(\D_{v_{(d)}}v_{(b)})-\D_{v_{(d)}}(\D_{v_{(c)}}v_{(b)})-\D_{[v_{(c)},v_{(d)}]}v_{(b)})\\
&=\g(v_{(a)},\D_{v_{(c)}}(\g^{(m)(n)}\Gamma_{(m)(b)(d)}v_{(n)}))-\g(v_{(a)},\D_{v_{(d)}}(\g^{(m)(n)}\Gamma_{(m)(b)(c)}v_{(n)}))\\
&+\g^{(m)(n)}\Gamma_{(a)(b)(n)}(\Gamma_{(m)(c)(d)}-\Gamma_{(m)(d)(c)})\\
&=v_{(c)}(\Gamma_{(a)(b)(d)})-v_{(d)}(\Gamma_{(a)(b)(c)})+\g^{(m)(n)}\Gamma_{(a)(b)(n)}(\Gamma_{(m)(c)(d)}-\Gamma_{(m)(d)(c)})\\
&+\g_{(a)(n)}[\Gamma_{(m)(b)(d)}v_{(c)}(\g^{(m)(n)})-\Gamma_{(m)(b)(c)}v_{(d)}(\g^{(m)(n)})]\\
&+\g^{(m)(n)}(\Gamma_{(m)(b)(d)}\Gamma_{(a)(n)(c)}-\Gamma_{(m)(b)(c)}\Gamma_{(a)(n)(d)}).
\end{split}
\end{equation*}
We set $d=4$ and use $\Gamma_{(a)(b)(4)}=v_{(4)}(\g^{(a)(b)})=0$ and $\g^{(a)(b)}=h^{(a)(b)}$; the result is
\begin{equation*}
\Lb(\Gamma_{(a)(b)(c)})=-h^{(m)(n)}\Gamma_{(a)(b)(n)}\Gamma_{(m)(4)(c)}-\R_{(a)(b)(c)(4)}.
\end{equation*}
Similarly,
\begin{equation*}
\Lb({\Gamma'}_{(a)(b)(c)})=-h^{(m)(n)}{\Gamma'}_{(a)(b)(n)}{\Gamma'}_{(m)(4)(c)}-{\R'}_{(a)(b)(c)(4)}.
\end{equation*}
We subtract these two identities to derive
\begin{equation}\label{a20}
\Lb[(d\Gamma)_{(a)(b)(c)})]={}^{(1)}F_{(a)(b)(c)}^{(d)(e)(f)}(d\Gamma)_{(d)(e)(f)}-(dR)_{(a)(b)(c)(4)}
\end{equation}
for some smooth function ${}^{(1)}F$. This can be written schematically in the form
\begin{equation}\label{schem1}
\Lb(d\Gamma)=\mathcal{M}_\infty(d\Gamma)+\mathcal{M}_\infty(d R).
\end{equation}
We will use such schematic equations for simplicity of notation\footnote{In general, given $B=(B_1,\ldots B_L):\O(x_0)\to\mathbb{R}^L$ we let $\mathcal{M}_\infty(B):\O(x_0)\to\mathbb{R}^{L'}$ denote vector-valued functions of the form ${\mathcal{M}_\infty(B)}_{l'}=\sum_{l=1}^LA_{l'}^lB_l$, where the coefficients $A_{l'}^l$ are smooth on $\O(x_0)$.}. 

For $a,b,c=1,\ldots,4$ and $\al=0,\ldots,3$ we define 
\begin{equation*}
\begin{split}
&(\partial d\Gamma)_{\al(a)(b)(c)}=\partial_\al[(d\Gamma)_{(a)(b)(c)}];\\
&(\partial dR)_{\al(a)(b)(c)(d)}=\partial_\al[(dR)_{(a)(b)(c)(d)}],
\end{split}
\end{equation*}
where $\partial_\al$ are the coordinate   vector-fields relative to our local coordinates
in $\O(x_0)$. By differentiating \eqref{schem1},
\begin{equation}\label{schem2}
\Lb(\partial  d\Gamma)=\mathcal{M}_\infty(d\Gamma)+\mathcal{M}_\infty(\partial d\Gamma)+\mathcal{M}_\infty(dR)+\mathcal{M}_\infty(\partial dR).
\end{equation}

Assume now that
\begin{equation*}
\begin{split}
&v_{(a)}=v_{(a)}^\al \pr_\al,\qquad{v'}_{(a)}={v'}_{(a)}^\al \pr_\al,\\
&{v'}_{(a)}-v_{(a)}=(dv)_{(a)}^{\al}\pr_\al,\quad (dv)_{(a)}^{\al}={v'}_{(a)}^\al-v_{(a)}^\al,
\end{split}
\end{equation*}
are the representations of the vectors $v_{(a)}$, ${v'}_{(a)}$, and ${v'}_{(a)}-v_{(a)}$ in our  coordinate frame $\{\pr_\al\}_{\al=0,\ldots,3}$. Since $[v_{(4)},v_{(b)}]=-\D_{v_{(b)}}v_{(4)}=-{\Gamma^{(c)}}_{(4)(b)}v_{(c)}$, we have
\begin{equation*}
v_{(4)}^\al\pr_\al(v_{(b)}^\be)-v_{(b)}^\al\pr_\al(v_{(4)}^\be)=-{\Gamma}_{(a)(4)(b)}v_{(c)}^\be\g^{(a)(c)}.
\end{equation*}
Similarly,
\begin{equation*}
v_{(4)}^\al\pr_\al({v'}_{(b)}^\be)-{v'}_{(b)}^\al\pr_\al(v_{(4)}^\be)=-{\Gamma'}_{(a)(4)(b)}{v'}_{(c)}^\be{\g'}^{(a)(c)}.
\end{equation*}
We subtract these two identities to conclude that, schematically,
\begin{equation}\label{schem3}
\Lb(dv)=\mathcal{M}_\infty(d\Gamma)+\mathcal{M}_\infty(dv).
\end{equation}
As before, we define
\begin{equation*}
(\partial dv)_{\al(b)}^\be=\pr_\al[(dv)_{(b)}^\be].
\end{equation*}
By differentiating \eqref{schem3} we have
\begin{equation}\label{schem4}
\Lb(\partial dv)=\mathcal{M}_\infty(d\Gamma)+\mathcal{M}_\infty(\partial d\Gamma)+\mathcal{M}_\infty(dv)+\mathcal{M}_\infty(\partial dv).
\end{equation}

Finally, we derive a wave equation for $dR$. We start from the identity
\begin{equation*}
(\square_\g  \R)_{(a)(b)(c)(d)}-(\square_{\g'}{\R'})_{(a)(b)(c)(d)}=\mathcal{M}_\infty(dR),
\end{equation*}
which follows from the standard wave equations satisfied by $\R$ and $\R'$ and the fact that $\g^{(m)(n)}={\g'}^{(m)(n)}=h^{(m)(n)}$. We also have
\begin{equation*}
\begin{split}
&\D_{(m)}\R_{(a)(b)(c)(d)}-{\D'}_{(m)}{\R'}_{(a)(b)(c)(d)}\\
&=\mathcal{M}_\infty(dv)+\mathcal{M}_\infty(d\Gamma)+\mathcal{M}_\infty(d\R)+\mathcal{M}_\infty(\partial d\R).
\end{split}
\end{equation*}
It follows from the last two equations that
\begin{equation*}
\begin{split}
&\g^{(m)(n)}v_{(n)}(v_{(m)}(\R_{(a)(b)(c)(d)}))-{\g'}^{(m)(n)}{v'}_{(n)}({v'}_{(m)}({\R'}_{(a)(b)(c)(d)}))\\
&=\mathcal{M}_\infty(dv)+\mathcal{M}_\infty(d\Gamma)+\mathcal{M}_\infty(\partial d\Gamma)+\mathcal{M}_\infty(dR)+\mathcal{M}_\infty(\partial dR).
\end{split}
\end{equation*}
Since $\g^{(m)(n)}={\g'}^{(m)(n)}$ it follows that
\begin{equation*}
\begin{split}
&\g^{(m)(n)}v_{(n)}(v_{(m)}((dR)_{(a)(b)(c)(d)}))\\
&=\mathcal{M}_\infty(dv)+\mathcal{M}_\infty(\partial dv)+\mathcal{M}_\infty(d\Gamma)+\mathcal{M}_\infty(\partial d\Gamma)+\mathcal{M}_\infty(dR)+\mathcal{M}_\infty(\partial dR).
\end{split}
\end{equation*}
Thus
\begin{equation}\label{schem5}
\square_\g(dR)=\mathcal{M}_\infty(dv)+\mathcal{M}_\infty(\partial dv)+\mathcal{M}_\infty(d\Gamma)+\mathcal{M}_\infty(\partial d\Gamma)+\mathcal{M}_\infty(dR)+\mathcal{M}_\infty(\partial dR).
\end{equation}
This is our main wave equation.

We collect now equations \eqref{schem1}, \eqref{schem2}, \eqref{schem3}, \eqref{schem4}, and \eqref{schem5}:
\begin{equation}\label{schem50}
\begin{split}
&\Lb(d\Gamma)=\mathcal{M}_\infty(d\Gamma)+\mathcal{M}_\infty(dR);\\
&\Lb(\partial d\Gamma)=\mathcal{M}_\infty(d\Gamma)+\mathcal{M}_\infty(\partial d\Gamma)+\mathcal{M}_\infty(dR)+\mathcal{M}_\infty(\partial dR);\\
&\Lb(dv)=\mathcal{M}_\infty(dv)+\mathcal{M}_\infty(d\Gamma);\\
&\Lb(\partial dv)=\mathcal{M}_\infty(dv)+\mathcal{M}_\infty(\partial dv)+\mathcal{M}_\infty(d\Gamma)+\mathcal{M}_\infty(\partial d\Gamma);\\
&\square_\g(dR)=\mathcal{M}_\infty(dv)+\mathcal{M}_\infty(\partial dv)+\mathcal{M}_\infty(d\Gamma)+\mathcal{M}_\infty(\partial d\Gamma)+\mathcal{M}_\infty(dR)+\mathcal{M}_\infty(\partial dR).
\end{split}
\end{equation}
This is our main system of equations. Since $\g=\g'$ in $I^{++}\cup I^{--}$, it follows easily that the functions $d\Gamma$, $\partial d\Gamma$, $dv$, $\partial dv$ and $dR$ vanish also  in $I^{++}\cup I^{--}$. Therefore, the proposition follows from Lemma \ref{extendedCarl} below.
\end{proof}

\begin{lemma}\label{extendedCarl}
Assume  $G_i,H_j:\O(x_0)\to\mathbb{R}$ are smooth functions, $i=1,\ldots,I$, $j=1,\ldots,J$. Let $G=(G_1,\ldots,G_I)$, $H=(H_1,\ldots,H_J)$, $\partial G=(\pr_0G_1,\ldots,\pr_4G_I)$ and assume that in $\O(x_0)$,
\begin{equation}\label{Haw60}
\begin{cases}
&\square_\g G=\mathcal{M}_\infty(G)+\mathcal{M}_\infty(\partial G)+\mathcal{M}_\infty(H);\\
&\Lb(H)=\mathcal{M}_\infty(G)+\mathcal{M}_\infty(\partial G)+\mathcal{M}_\infty(H).
\end{cases}
\end{equation}
Assume that $G=0$ and $H=0$ on $(\NN\cup\underline{\NN})\cap \O(x_0)$. Then,
 there exists a small neighborhood $\O'(x_0)\subset\O(x_0)$ of $x_0$ such that $G=0$ and $H=0$ in $\big(I^{+-} \cup I^{-+}\big)\cap \O'(x_0) $.
\end{lemma} 
Unique continuation theorems of this type in the case $H=0$ were proved by two of the authors in \cite{IoKl} and \cite{IoKl2}, using Carleman estimates. It is not hard to adapt the proofs, using similar Carleman estimates, to the general case; we provide all the details in the appendix. This completes the proof of Theorem \ref{mainfirststep}.

We show now that the Killing vector-field $\K$ is timelike, in a quantitative sense, in a small neighborhood of $S$ in the complement of $I^{++}\cup I^{--}$.

\begin{proposition}\label{Ktimelike}
 Let  $\K$ be the  Killing vector-field, constructed above,  in a neighborhood  $\O$ of $S$.  Then there is a neighborhood $\O'\subset\O$ of $S$ such that
\begin{equation}\label{Haw65}
\g(\K,\K)\leq u\underline{u}\quad\text{ in }(I^{+-}\cup I^{-+})\cap\O'.
\end{equation}
In particular, the vector-field $\K$ is timelike in the set $\O'\setminus (I^{++}\cup I^{--})$.
\end{proposition}

\begin{proof}[Proof of Proposition \ref{Ktimelike}] Since $\K$ is a Killing vector-field in $\O$, we have
\begin{equation}\label{Haw66}
\square_\g(\K^\be\K_\be)=2\D^\al(\K^\be\D_\al\K_\be)=2\D^\al\K^\be\D_\al\K_\be=-4\quad\text{ on }S.
\end{equation}
Indeed, $\square_\g \K=0$ and it follows from \eqref{Haw17} that $2\D^\al\K^\be\D_\al\K_\be=4\D^3\K^4\D_3\K_4=-4$ on $S$. Since $\K_\be\K^\be=0$ on $(\NN\cup\underline{\NN})\cap\O$ (see \eqref{Haw5}), we have $\K_\be\K^\be=u\underline{u}f$ on $\O$ for some smooth function $f:\O\to\mathbb{R}$. Using \eqref{Haw66} on $S$ and the fact that $u=\underline{u}=0$ on $S$, we derive
\begin{equation*}
-4=\D^\al\D_\al(u\underline{u}f)=2f\D^\al u\D_\al\underline{u}=-2f\Lb(u)L(\underline{u})=-2f.
\end{equation*}
Thus $f=2$ on $S$, and the bound \eqref{Haw65} follows for  a  sufficiently small $\O'$.
\end{proof}

\section{Further results in the presence of a symmetry}\label{additional}

The goal of this section is to prove Theorem \ref{thm:rotation}. So far  we have constructed  a smooth Killing vector-field $\K$ defined in an  open set $\O$ such that $\K=\underline{u}L-u\underline{L}$ on $(\NN\cup\underline{\NN})\cap\O$.

Assume in this section that the space-time $(\O,\g)$ admits another  smooth Killing vector-field $\T$, which is tangent to the null hypersurfaces $\NN$ and $\underline{\NN}$. We recall several definitions (see \cite[Section 4]{IoKl} for a longer discussion and proofs of some identities). In $\O$ we define the 2-form $F_{\al\be}=\D_\al\T_\be$ and the complex valued 2-form,
\begin{equation}\label{s1}
\FF_{\al\be}=F_{\al\be}+i {\dual F}_{\al\be}=F_{\al\be}+(i/2){\in_{\al\be}}^{\mu\nu}F_{\mu\nu}.
\end{equation}
Let $\FF^2=\FF_{\al\be}\FF^{\al\be}$. We define also the Ernst $1$-form
\begin{equation}\label{s12}
\si_\mu=2\T^\a\FF_{\a\mu}=\D_\mu(-\T^\al \T_\al)- i\in_{\mu\b\ga\de}\T^\b \D^\ga\T^\de.
\end{equation}
It is easy to check that, in $\O$
\begin{equation}\label{Ernst1}
\begin{cases}
&\D_\mu\si_\nu-\D_\nu\si_\mu=0;\\
&\D^\mu\si_\mu=-\FF^2;\\
&\si_\mu\si^\mu=\g(\T,\T) \FF^2.
\end{cases}
\end{equation}

\begin{proposition}\label{TKcommute}
There is an open set $\O'\subseteq \O$, $S\subseteq \O'$ such that
\begin{equation}\label{haw201}
[\T,\K]=0\text{ in }\O'.
\end{equation}
In addition, if $\si_\mu=2\T^\al\FF_{\al\mu}$ is the Ernst $1$-form associated to $\T$ (see \eqref{s12}), then
\begin{equation}\label{haw200}
\K^\mu\si_\mu=0\text{ in }\O'.
\end{equation}
\end{proposition}

\begin{proof}[Proof of Proposition \ref{TKcommute}] We show first that 
\begin{equation}\label{Haw55}
[\T,\K]=0\quad\text{ on }(\NN\cup\underline{\NN})\cap\O.
\end{equation}
By symmetry, it suffices  to check that $[\T,\K]=0$ on $\NN\cap\O$. We first observe  that $[\T, L]$ is proportional to $L$.
 Indeed, since  the null second fundamental form of $\NN$ is symmetric  and $\T$ is  both  Killing and tangent to $\NN$, we have 
  for every $X\in T(\NN)$,
 \beaa
   \g([\T, L], X)&=&\g(\D_\T L, X)-\g(\D_L \T, X)= \g(\D_\T L, X)  +\g(\D_X \T, L)\\
  &=&    \g(\D_\T L, X)-\g(\T, \D_X L)=\chi(\T, X)-\chi(X, \T)=0.
  \eeaa
Consequently   $[\T, L]$ must be proportional to $L$, i.e. $[\T, L]=f L$.
Since $\D_L L=0$  and $\T$ commutes with covariant derivatives we 
derive,
 \beaa
0&=& \Lie_\T( \D_ L L)= \D_{\Lie_T L}L+\D_{L}(\Lie_\T L)\\
&=&\D_{ fL}L+\D_{L}(f L)=L( f) L.
 \eeaa
Therefore
\begin{equation}\label{hw21}
[\T, L]=f L\qquad\text{ and }\qquad L(f)=0\qquad\text{ on }\NN\cap\O.
\end{equation}

On the other hand, in view of the definition 
 of $\underline{u}$ we have $\T(L(\underline{u}))-L(\T (\underline{u}))=f L(\underline{u})$. Hence,
 \beaa
 L( f \underline{u}+\T (\underline{u}))=0.
 \eeaa
 Since  $\T$ is tangent  to $S$  and $\underline{u}=0$ on $S$, we deduce that
 $f \underline{u}+\T (\underline{u})$ vanishes on $S$, thus
  \beaa
   \T \underline{u}+f \underline{u}=0,\qquad \mbox{on}\,\,\, \NN\cap\O.
  \eeaa
 Now, $[\T,\underline{u}L]=\T(\underline{u})L+\underline{u}[\T, L]=\big(\T(\underline{u})+f\underline{u} \big)L=0$. The identity  \eqref{Haw55} follows since $\K=\underline{u}L$ on $\NN\cap\O$.

Let $V=[\T,\K]=\mathcal{L}_{\T}\K$ on $\O$. Since $\square_\g\K=0$ and $\T$ is Killing, we derive, after commuting covariant and Lie derivatives,
\begin{equation*}
0=\mathcal{L}_{\T}(\square_\g\K)=\square_\g(\mathcal{L}_{\T}\K)=\square_\g V.
\end{equation*}
Since $V$ vanishes on $(\NN\cup\underline{\NN})\cap\O$, it follows that $V$ vanishes in $(I^{++}\cup I^{--})\cap\O'$, for some smaller  neighborhood  $\O'$ of $S$. (due to the well-posedness of the characteristic initial-value problem); it also follows that $V$ vanishes in $(I^{+-}\cup I^{-+})\cap\O'$ using Lemma \ref{extendedCarl} with $H=0$. This completes the proof of \eqref{haw201}. 

We prove now the identity \eqref{haw200}. Since $\K$ and $\T$ commute  we  observe that
 $\Lie_\K\FF=0$ in $\O$. 
 In addition, since $\square_\g \K=0$, $\D \K$ is antisymmetric, 
$\D\si$ is symmetric with trace  $\D^\al\si_\al=-\FF^2$     (see \eqref{Ernst1}) and $\mbox{\bf Ric}(\g)=0$,
we  have  in $\O$
\begin{equation}\label{haw207}
\square_\g(\K^\mu\si_\mu)=\K^\mu \square_\g \si_\mu=
\K^\mu\D_\mu(\D^\al\si_\al)=-\mathcal{L}_{\K}\FF^2=0.
\end{equation}
We show below that  the function $\K^\mu\si_\mu$ vanishes on $(\NN\cup\underline{\NN})\cap\O$. Thus,  as before, we conclude that $\K^\mu\si_\mu=0$ in a smaler neighborhood  $\O'$, as desired. 

To show that $\K^\mu\si_\mu$ vanishes on $(\NN\cup\underline{\NN})\cap\O$ we calculate with respect
to our null  frame $L=e_4$, $\Lb=e_3$, $e_1$, $e_2$  defined  in a neighborhood of $S$. Since $\T$ is  tangent to $\NN$, for $a=1,2$ we have
$
F_{a4}=e_a(\g(\T,e_4))-\g(\T,\D_{e_a}e_4)=0
$
along $\NN$ (since $\D_{e_a}e_4=-\ze_a e_4$, see \eqref{table1}). Similarly, $F_{a3}=0$ along $\underline{\NN}$. Thus
\begin{equation}\label{Haw205}
\FF_{14}=\FF_{24}=0\text{ on }\NN\cap\O\quad\text{ and }\quad\FF_{13}=\FF_{23}=0\text{ on }\underline{\NN}\cap\O.
\end{equation}
Since $\K=\underline{u}e_4-ue_3$ on $(\NN\cup\underline{\NN})\cap\O$, we infer that,
\begin{equation}\label{haw206}
\K^\mu\si_\mu=2\K^\mu\T^\al\FF_{\al\mu}=0\quad\text{ on }(\NN\cup\underline{\NN})\cap\O,
\end{equation}
as desired.
\end{proof}

\begin{proposition}\label{goodcuts} There is a constant $\lambda_0\in\mathbb{R}$ and an open neighborhood $\O'\subseteq\O$ of $S$ such that the vector-field
\begin{equation*}
\Z=\T+\lambda_0\K
\end{equation*}
has periodic  orbits in $\O'$. In other words , there is $t_0>0$ such that $\Psi_{t_0,\Z}=\mathrm{Id}$ in $\O'$.
\end{proposition}
This completes the proof of Theorem \ref{thm:rotation}.
Observe that the  main constants $\lambda_0$ and $t_0$ can be determined on the bifurcation sphere $S$. We show below  that Proposition \ref{goodcuts} follows from the following lemma.
\begin{lemma}\label{Sdef}
There is a constant $t_0>0$ such that $\Psi_{t_0,\T}=\mathrm{Id}$ in $S$. In addition, there is a constant $\lambda_0\in\mathbb{R}$ and a choice of the null pair $(L,\underline{L})$ along $S$ (satisfying \eqref{normalization}) such that
\begin{equation}\label{hw20}
[\T,L]=\lambda_0L\quad\text{ and }\quad [\T,\underline{L}]=-\lambda_0\underline{L}\quad\text{ on }S.
\end{equation}
\end{lemma}
\begin{proof}[Proof of Proposition \ref{goodcuts}] It follows from \eqref{hw21} and \eqref{hw20} that
\begin{equation}\label{hw001}
[\T,L]=\lambda_0L\quad\text{ on }\NN\cap\O\quad\text{ and }\quad[\T,\underline{L}]=-\lambda_0\underline{L}\quad\text{ on }\underline{\NN}\cap\O.
\end{equation}
Thus, using the identity $[\underline{L},\K]=-\underline{L}$ in Proposition \ref{Friedrich},
\begin{equation*}
[\Z,\underline{L}]=[\T+\lambda_0\K,\underline{L}]=0\quad\text{ on }\underline{\NN}\cap\O.
\end{equation*}
Since $\Z$ is a Killing vector-field, it follows as in the proof of Proposition \ref{Friedrich} (see \eqref{eq:Lie.ident}) that 
\begin{equation*}
[\Z,\underline{L}]=0\quad\text{ in }\O.
\end{equation*}

An argument similar to the proof of \eqref{hw00} shows that $[L,\K]-L=0$ on $\underline{\NN}\cap\O$. Using the first identity in \eqref{hw001}, it follows that $[\Z,L]=0$ on $\underline{\NN}\cap\O$. Since $\Z$ is a Killing vector-field, it follows as in Proposition \ref{Friedrich} that $[\Z,L]=0$ in $\O$. 

The conclusion of the proposition follows from the first claim in  Lemma \ref{Sdef} and the identities $[\Z,\underline{L}]=[\Z,L]=0$ in $\O$.  
\end{proof}

\begin{proof}[Proof of Lemma \ref{Sdef}] The existence of the period $t_0$ is a standard fact concerning Killing vector-fields on the sphere\footnote{If $\T\equiv 0$ on $S$ then any value of $t_0>0$ is suitable. In this case, the conclusion of Proposition \ref{goodcuts} is that $\T+\lambda_0\K\equiv 0$ in $\O'$ for some $\lambda_0\in\mathbb{R}$.}.
In particular all  nontrivial  orbits of $S$ are compact and diffeomorphic to $\mathbb{S}^1$. To prove \eqref{hw20}, in view of \eqref{hw21} it suffices to prove that there is $\lambda_0\in\mathbb{R}$ and a choice of the null pair $(L,\Lb)$ on $S$ such that
\begin{equation*}
\g([\T,L],\Lb)=-\lambda_0,\qquad \g([\T,\Lb],L)=\lambda_0\qquad\text{ on }S.
\end{equation*}
Both identities are equivalent to 
\begin{equation*}
\T^\al\Lb^\be\D_\al L_\be-L^\al\Lb^\be\D_\al\T_\be=-\lambda_0, 
\end{equation*}
which is equivalent to
   \beaa
 \la_0=F_{43}-\g(\ze, \T).
   \eeaa
   We thus have to show that there exist a choice of the null pair
   $e_4=\Lp, e_3=\Lm$ along $S$ such that the scalar function below is constant  along $S$,
   \bea
 H:=  F_{43}-\g(\ze, \T).\label{def:H}
   \eea
   Under   a scaling  transformation
   $e_4'=f e_4, e_3'=f^{-1} e_3$ the torsion $\ze$  changes according to the formula, 
   \beaa
   \ze'=\ze-\nabla \log f.
   \eeaa
   Therefore, in the new frame,
   \beaa
   H'=F_{4'3'}-\g(\ze', \T)= F_{43}-\g(\ze, \T)+\T(\log f)=H+\T(\log f)
   \eeaa
   Consequently, we are led  to look for a function $f$ such that $H+\T(\log f)$
   is a constant.  Taking $\hat H$ to be the average of $H$ along the integral curves of $\T$  and solving the equation 
   \bea
   \T(\log f)&=&-H+\hat H,
   \eea
    it  only remains to prove that $\hat H$ is constant along $S$.
    
 Since $\T$ is Killing we must have, 
\bea
\D_\a \D_\b T_\ga=T^\la \R_{\la\a\b\ga}\label{Kill-curv}
\eea
  Using \eqref{Kill-curv}  and the formulas  \eqref{table1} on $S$  we derive,
    \beaa
\T^\la \R_{\la a43}   &=&  \D_a\D_4 \T_3=e_a(\D_4 \T_3)=e_a(F_{43}).
    \eeaa
    Thus,  since $\T$ is tangent to $S$ and   $ \T^b \R_{b a43}=\frac 1 2 \in_{ab}\T^b\, \si $
    (with $\si= \dual\R_{3434}$)
    \bea
    e_a(F_{43})&=&\T^b \R_{b a43}=\frac 1 2 \in_{ab}\T^b\, \si.
    \eea
    In particular, the function  $H$ defined in \eqref{def:H} is constant on $S$ if $\T\equiv 0$ on $S$. Thus we may assume in the rest of the proof that the set $\Lambda=\{p\in S:\T_p=0\}$ is finite.
    
    On the other hand, writing $\nabla_a\ze_b-\nabla_b\ze_a=\in_{ab}\curl \ze$,
    \beaa
    e_a\g(\ze, \T)&=&\nabla_a \ze_b \T^b+ \ze_b \nabla_a \T_b =(\nabla_a \ze_b-\nabla_b\ze_a) \T^b+ \ze^b \nabla_a \T_b+\nabla_\T \ze_a\\
    &=& \in_{ab} \curl\ze\T^b+ \ze^b \nabla_a \T_b+\nabla_\T \ze_a
    \eeaa
    The torsion $\ze$ verifies the equation, 
    \bea
    \curl \ze=\frac 1 2 \si,
    \eea
    Therefore,
     \bea
    e_a\g(\ze, \T)&=&\frac 1 2 \in_{ab}\T^b\, \si + \ze^b \nabla_a \T_b+\nabla_\T \ze_a.
   \eea
    Since $H=F_{43}-\ze\c\T$    we deduce,
    \bea
    e_a(H)&=&- \ze^b \nabla_a \T_b-\nabla_\T \ze_a.
    \eea
    
    Consider   the orthonormal frame $e_1, e_2$ on $S\setminus\Lambda$,
   \beaa
   e_1=X^{-1} \T,\qquad X^2=\g(\T, \T).
   \eeaa
 Since $e_1(X)=0$ and $e_1=X^{-1}\T$, we have 
   \beaa
   \nabla_\T e_2=- F_{12} e_1.
   \eeaa
   We claim that, with respect to this local  frame, 
   \bea
   \nabla_2(H)=-\T(\ze_2).\label{formula.crucial}
   \eea
   Indeed,
   \beaa
   \nabla_2 (H)&=&-\ze^1 \nabla_2\T_1-\ze^2\nabla_2 \T_2-\g(\nabla_\T \ze, e_2)\\
   &=&-\ze^1 F_{21} -\T\g(\ze, e_2)+\g(\ze, \nabla_\T e_2)\\
   &=&-\T\g(\ze, e_2)-\ze^1  F_{21}-\ze^1 F_{12}\\
   &=&-\T(\ze_2)
   \eeaa
   We now fix a a non-trivial  orbit $\ga_0$   of $\T$ in $S\setminus\Lambda$. Consider the geodesics initiating on  $\ga_0$ and perpendicular to it and $\phi$ the corresponding affine parameter.  More precisely we choose  a vector  $V$ on $\ga_0$ such that $\g(V,V)=1$ and extend it by parallel transport along the geodesics perpendicular
   to $\ga_0$. Then choose $\phi$ such that $V(\phi)=1$ and $\phi=0$  on $\ga_0$.
   This defines a system of coordinates $t, \phi$ in a neighborhood  $U$  of $\ga_0$, such that $\pr_t=T$,   $\nabla_{\pr_\phi} \pr_\phi=0$ in   $U$  and $\g(\pr_t,\pr_\phi)=0$,
    $\g(\pr_\phi, \pr_\phi)=1$  on  $\Ga_0$.
   Since $\pr_t $ is Killing we must have $X^2=-\g(\pr_t, \pr_t)$  and $\g(\pr_\phi,\pr_\phi)$ independent of $t$. Moreover,
   \beaa
   \pr_\phi\g(\pr_t,\pr_\phi)&=&\g(\nabla_{\pr_\phi }\pr_t,\pr_\phi)+\g( \pr_t, \nabla_{\pr_\phi} \pr_\phi)=\g(\nabla_{\pr_t }\pr_\phi,\pr_\phi)
   =\frac 1 2 \pr_t\g(\pr_\phi,\pr_\phi)=0.
   \eeaa
   Hence, since $\g(\pr_t,\pr_\phi)=0$ on $\Ga_0$ we infer that  $\g(\pr_t,\pr_\phi)=0$
   in $U$. Similarly,
   \beaa
   \pr_\phi\g(\pr_\phi,\pr_\phi)=2\g(\nabla_{\pr_\phi}\pr_\phi,\pr_\phi)=0
   \eeaa
   and therefore, $\g(\pr_\phi,\pr_\phi)=1$ in $U$. Thus, in $U$,  the metric  $\g$ 
   takes the form,
   \bea
   d\phi^2+X^2(\phi) dt^2
   \eea   
  Therefore, with  $\T=\pr_t$,  $e_2=\pr_\phi$, 
  we deduce from \eqref{formula.crucial}, everywhere in $U$,
   \bea
   \pr_\phi H=-\pr_t\g(\ze, \pr_\phi)
   \eea
   Thus, integrating  in $t$  and in view of the fact that the 
   orbits  of $\pr_t$  are closed,   we infer that $\hat{H}$ is constant along $S$, as desired.
\end{proof}

\appendix

\section{Proof of Lemma \ref{extendedCarl}}
We will use a Carleman estimate proved by two of the authors in \cite[Section 3]{IoKl}, which we recall below.
Let $\O(x_0)$ a  coordinate neighborhood of a point $x_0\in S$ and   coordinates
$x^\a$ as in \eqref{haw1}.  We denote by $B_r=B_r(x_0)$,  the set 
of points  $p\in \O(x_0)$ whose coordinates $x=x^\a$ verify $|x-x_0|\le r$, relative  to the standard euclidean norm in $\O(x_0)$. 
Consider  two vector-fields  $V=V^\al\partial_\al,W=W^\al\partial_\al$  on $\O(x_0)$ which verify, that, 
\begin{equation}\label{po1}
\sup_{x\in \O(x_0)}\sum_{j=0}^4 (|\partial^jV(x)|+|\partial^jW(x)|)
\leq A,
\end{equation}
where $A$ is a  large constant (as in \eqref{haw1}), and $ |\partial^jV(x)| $ denotes the sum of the absolute values of all partial derivatives 
of order $j$ of all components of $V$ in our given coordinate  system. When $j=1$ we write simply $ |\partial V(x)| $.

\begin{definition}\label{psconvex}
A family of weights $h_\eps:B_{\eps^{10}}\to\mathbb{R}_+$, $\eps\in(0,\eps_1)$, $\eps_1\leq A^{-1}$, will be called $V$-conditional pseudo-convex if for any $\eps\in(0,\eps_1)$ \begin{equation}\label{po5}
\begin{split}
h_\eps(x_0)=\eps,\quad\sup_{x\in B_{\eps^{10}}}\sum_{j=1}^4\eps^j|\partial^jh_\eps(x)|\leq\eps/\eps_1,\quad |V(h_\eps)(x_0)|\leq\eps^{10},
\end{split}
\end{equation}
\begin{equation}\label{po3.2}
\D^\alpha h_\eps(x_0)\D^\be h_\eps(x_0)(\D_\al h_\eps\D_\be h_\eps-\eps\D_\al\D_\be h_\eps)(x_0)\geq\eps_1^2,
\end{equation}
and there is $\mu\in[-\eps_1^{-1},\eps_1^{-1}]$ such that for all vectors $X=X^\alpha\partial_\alpha$ at $x_0$
\begin{equation}\label{po3}
\begin{split}
&\eps_1^2[(X^1)^2+(X^2)^2+(X^3)^2+(X^4)^2]\\
&\leq X^\al X^\be(\mu\g_{\al\be}-\D_\al\D_\be h_\eps)(x_0)+\eps^{-2}(|X^\al V_\al(x_0)|^2+|X^\al\D_\al h_\eps(x_0)|^2).
\end{split}
\end{equation}
A function $e_\eps:B_{\eps^{10}}\to\mathbb{R}$ will be called a negligible perturbation if
\begin{equation}\label{smallweight}
\sup_{x\in B_{\eps^{10}}}|\partial^je_\eps(x)|\leq\eps^{10}\qquad\text{ for  }j=0,\ldots,4.
\end{equation}
\end{definition}

Our main Carleman estimate, see \cite[Section 3]{IoKl}, is the following:

\begin{lemma}\label{Cargen}
Assume $\eps_1\leq A^{-1}$, $\{h_\eps\}_{\eps\in(0,\eps_1)}$ is a $V$-conditional pseudo-convex family, and $e_\eps$ is a negligible perturbation for any $\eps\in(0,\eps_1]$. Then there is $\eps\in (0,\eps_1)$ sufficiently small (depending only on $\eps_1$) and $\widetilde{C}_\eps$ sufficiently large such that for any $\lambda\geq\widetilde{C}_\eps$ and any $\phi\in C^\infty_0(B_{\eps^{10}})$
\begin{equation}\label{Car1gen}
\lambda \|e^{-\lambda f_\eps}\phi\|_{L^2}+\|e^{-\lambda f_\eps}|\partial \phi|\,\|_{L^2}\leq \widetilde{C}_\eps\lambda^{-1/2}\|e^{-\lambda f_\eps}\,\square_{\g}\phi\|_{L^2}+\eps^{-6}\|e^{-\lambda f_\eps}V(\phi)\|_{L^2},
\end{equation}
where $f_\ep=\ln (h_\eps+e_\eps)$.
\end{lemma}

We will only use this Carleman estimate with $V=0$. In this case the pseudo-convexity condition in Definition \ref{psconvex} is a special case of H\"{o}rmander's pseudo-convexity condition \cite[Chapter 28]{Ho}. We also need a Carleman estimate to exploit the ODE's in \eqref{Haw60}.

\begin{lemma}\label{CarODE}
Assume $\eps\leq A^{-1}$ is sufficiently small, $e_\eps$ is a negligible perturbation, and $h_\eps:B_{\eps^{10}}\to\R_+$ satisfies
\begin{equation}\label{ODEneeds}
h_{\eps}(x_0)=\eps,\quad\sup_{x\in B_{\eps^{10}}}\sum_{j=1}^2\eps^j|\partial^jh_{\eps}(x)|\leq 1,\quad |W(h_\eps)(x_0)|\geq 1.
\end{equation}
Then there is $\widetilde{C}_\eps$ sufficiently large such that for any $\lambda\geq\widetilde{C}_\eps$ and any $\phi\in C^\infty_0(B_{\eps^{10}})$
\begin{equation}\label{Car1ODE}
\|e^{-\lambda f_\eps}\phi\|_{L^2}\leq 4\lambda^{-1}\|e^{-\lambda f_\eps}W(\phi)\|_{L^2},
\end{equation}
where $f_\ep=\ln (h_\eps+e_\eps)$.
\end{lemma} 

\begin{proof}[Proof of Lemma \ref{CarODE}] Clearly, we may assume that $\phi$ is real-valued and let $\psi=e^{-\lambda f_\eps}\phi\in C^\infty_0(B_{\eps^{10}})$. We have to prove that
\begin{equation}\label{Car2ODE}
\|\psi\|_{L^2}\leq 4\|\lambda^{-1}W(\psi)+W(f_\eps)\psi\|_{L^2}.
\end{equation}
By integration by parts,
\begin{equation*}
\begin{split}
&\int_{B_{\eps^{10}}}[\lambda^{-1}W(\psi)+W(f_\eps)\psi]\cdot W(f_\eps)\psi\,d\mu\\
&=\int_{B_{\eps^{10}}}[W(f_\eps)\psi]^2\,d\mu-(2\lambda)^{-1}\int_{B_{\eps^{10}}}\psi^2\cdot\D_\al(W(f_\eps)W^\al)d\mu.
\end{split}
\end{equation*}
In view of \eqref{ODEneeds} and the assumption \eqref{po1}
\begin{equation*}
|W(f_\eps)|\geq 1\quad\text{ and }\quad|\D_\al(W(f_\eps)W^\al)|\leq\widetilde{C}_\eps\quad\text{ in }\quad B_{\eps^{10}},
\end{equation*}
provided that $\eps$ is sufficiently small. Thus, for $\lambda$ sufficiently large,
\begin{equation*}
\int_{B_{\eps^{10}}}[\lambda^{-1}W(\psi)+W(f_\eps)\psi]\cdot W(f_\eps)\psi\,d\mu\geq\frac{1}{2}\int_{B_{\eps^{10}}}[W(f_\eps)\psi]^2\,d\mu,
\end{equation*}
and the bound \eqref{Car2ODE} follows.
\end{proof}

\begin{proof}[Proof of Lemma \ref{extendedCarl}] It suffices to prove that $G=0$ and $H=0$ in $I_{\widetilde{c}}^{+-}$, for some $\widetilde{c}$ sufficiently small. We fix $x_0\in S$ and set
\begin{equation}\label{mw1}
h_\eps=\eps^{-1}(u+\eps)(-\underline{u}+\eps)\quad\text{and }\quad e_\eps=\eps^{10}N^{x_0},
\end{equation}
where $u,\underline{u}$ are the optical functions defined in section \ref{prelim} and $N^{x_0}(x)=|x- x_0|^2=\sum_{\a=0,1,2,3}|x^\a-x_0^\a|^2 $, the square of the standard euclidean norm. 

It is clear that $e_\eps$ is a negligible perturbation, in the sense of \eqref{smallweight}, for $\eps$ sufficiently small. Also, it is clear that $h_\eps$ verifies the condition \eqref{ODEneeds}, for $\eps$ sufficiently small and $W=2\underline{L}$. 

We show now that there is $\eps_1=\eps_1(A)$ sufficiently small such that the family of weights $\{h_\eps\}_{\eps\in(0,\eps_1)}$ is $0$-conditional pseudo-convex, in the sense of Definition \ref{psconvex}. Condition \eqref{po5} is clearly satisfied, in view of the definition and \eqref{haw10}. To verify conditions \eqref{po3.2} and \eqref{po3}, we compute, in the frame $e_1,e_2,e_3,e_4$ defined in section \ref{prelim},
\begin{equation}\label{mw2}
e_1(h_\eps)=e_2(h_\eps)=0,\quad e_3(h_\eps)=-\Omega(1-\eps^{-1}\underline{u}),\quad e_4(h_\eps)=\Omega(1+\eps^{-1}u)
\end{equation}
in $B_{\eps^{10}}(x_0)$, and
\begin{equation}\label{mw3}
\begin{split}
&(\D^2h_\eps)_{ab}=O(1),\quad (\D^2h_\eps)_{3a}=O(1),\quad (\D^2h_\eps)_{4a}=O(1),\quad a,b=1,2,\\
&(\D^2h_\eps)_{33}=O(1),\quad (\D^2h_\eps)_{44}=O(1),\quad (\D^2h_\eps)_{34}=-\Omega^2\eps^{-1}+O(1)
\end{split}
\end{equation}
in $B_{\eps^{10}}(x_0)$, where $O(1)$ denotes various functions on $B_{\eps^{10}}(x_0)$ with absolute value bounded by constants that depends only on $A$. Thus
\begin{equation*}
\D^\alpha h_\eps(x_0)\D^\be h_\eps(x_0)(\D_\al h_\eps\D_\be h_\eps-\eps\D_\al\D_\be h_\eps)(x_0)=2+\eps O(1).
\end{equation*}
This  proves \eqref{po3.2} if $\eps_1$ is sufficiently small. Similarly, if $X=X^\al e_\al$ then, with $\mu=\eps_1^{-1/2}$ we compute
\begin{equation*}
\begin{split}
&X^\al X^\be(\mu\g_{\al\be}-\D_\al\D_\be h_\eps)(x_0)+\eps^{-2}|X^\al\D_\al h_\eps(x_0)|^2\\
&=\mu((X^1)^2+(X^2)^2)+2(\eps^{-1}-\mu)X^3X^4+\eps^{-2}(X^3-X^4)^2+O(1)\sum_{\al=1}^4(X^\al)^2\\
&\geq (\mu/2)((X^1)^2+(X^2)^2)+(\eps^{-1}/2)((X^3)^2+(X^4)^2),
\end{split}
\end{equation*}
provided that $\eps_1$ is sufficiently small. This completes the proof of \eqref{po3}.

It follows from the Carleman estimates in Lemmas \ref{Cargen} and \ref{CarODE} that there is $\eps=\eps(A)\in(0,c)$ (where $c$ is the constant in Lemma \ref{extendedCarl}) and a constant $\widetilde{C}=\widetilde{C}(A)\geq 1$ such that
\begin{equation}\label{Bar1}
\begin{split}
&\lambda \|e^{-\lambda f_\eps}\phi\|_{L^2}+\|e^{-\lambda f_\eps}|\partial\phi|\,\|_{L^2}\leq \widetilde{C}\lambda^{-1/2}\|e^{-\lambda f_\eps}\,\square_{\g}\phi\|_{L^2};\\
&\|e^{-\lambda f_\eps}\phi\|_{L^2}\leq \widetilde{C}\lambda^{-1}\|e^{-\lambda f_\eps}\underline{L}(\phi)\|_{L^2},
\end{split}
\end{equation}
for any $\phi\in C^\infty_0(B_{\eps^{10}}(x_0))$ and any $\lambda\geq\widetilde{C}$, where $f_\ep=\ln (h_\eps+e_\eps)$. Let $\eta:\mathbb{R}\to[0,1]$ denote a smooth function supported in $[1/2,\infty)$ and equal to $1$ in $[3/4,\infty)$. For $\delta \in(0,1]$, $i=1,\ldots,I$, $j=1,\ldots J$ we define,
\begin{equation}\label{pr2}
\begin{split}
&G^{\delta,\eps}_i=G_i\cdot \mathbf{1}_{I_{c}^{+-}}\cdot \eta(-u\underline{u}/ \delta)\cdot \big(1-\eta(N^{x_0}/\eps^{20})\big)=G_i\cdot \widetilde{\eta}_{\delta,\eps}\\
&H^{\delta,\eps}_j=H_j\cdot \mathbf{1}_{I_{c}^{+-}}\cdot \eta(-u\underline{u}/ \delta)\cdot \big(1-\eta(N^{x_0}/\eps^{20})\big)=H_j\cdot \widetilde{\eta}_{\delta,\eps}.
\end{split}
\end{equation}
Clearly, $G^{\delta,\eps}_i,H^{\delta,\eps}_j\in C^\infty _0(B_{\eps^{10}}(x_0)\cap\mathbf{E})$. We would like to apply the inequalities in \eqref{Bar1} to the functions $G^{\delta,\eps}_i,H^{\delta,\eps}_j$, and then let $\delta \to 0$ and $\lambda \to\infty$ (in this order).

Using the definition \eqref{pr2}, we have
\begin{equation*}
\begin{split}
&\square_\g G^{\delta,\eps}_i=\widetilde{\eta}_{\delta,\eps}\cdot\square_\g G_i+2\D_\al G_i\cdot \D^\al \widetilde{\eta}_{\delta,\eps}+G_i\cdot \square_\g\widetilde{\eta}_{\delta,\eps};\\
&\underline{L}(H^{\delta,\eps}_j)=\widetilde{\eta}_{\delta,\eps}\cdot\underline{L}(H_j)+H_j\cdot\underline{L}(\widetilde{\eta}_{\delta,\eps}).
\end{split}
\end{equation*}
Using the Carleman inequalities \eqref{Bar1}, for any $i=1,\ldots,I$, $j=1,\ldots,J$ we have
\begin{equation}\label{va10}
\begin{split}
&\lambda\cdot \|e^{-\lambda f_{\eps}}\cdot \widetilde{\eta}_{\delta,\eps}G_i\|_{L^2}+\|e^{-\lambda f_{\eps}}\cdot \widetilde{\eta}_{\delta,\eps}|\partial^1G_i|\, \|_{L^2}\leq \widetilde{C}\lambda ^{-1/2}\cdot \|e^{-\lambda f_{\eps}}\cdot \widetilde{\eta}_{\delta,\eps}\square_\g G_i\|_{L^2}\\
&+\widetilde{C}\Big[\|e^{-\lambda f_{\eps}}\cdot \D_\al G_i\D^\al \widetilde{\eta}_{\delta,\eps} \|_{L^2}+\|e^{-\lambda f_{\eps}}\cdot G_i( |\square_\g\widetilde{\eta}_{\delta,\eps}|+|\partial^1\widetilde{\eta}_{\delta,\eps}| )\|_{L^2}\Big]
\end{split}
\end{equation}
and
\begin{equation}\label{va10.1}
\|e^{-\lambda f_{\eps}}\cdot \widetilde{\eta}_{\delta,\eps}H_j\|_{L^2}\leq \widetilde{C}\lambda^{-1}\|e^{-\lambda f_{\eps}}\cdot \widetilde{\eta}_{\delta,\eps}\underline{L}(H_j)\|_{L^2}+\widetilde{C}\lambda^{-1}\|e^{-\lambda f_{\eps}}\cdot H_j\underline{L}(\widetilde{\eta}_{\delta,\eps})\|_{L^2},
\end{equation}
for any $\lambda\geq\widetilde{C}$. Using the main identities \eqref{Haw60}, in $B_{\eps^{10}}(x_0)$ we estimate pointwise
\begin{equation}\label{va11}
\begin{split}
&|\square_\g G_i|\leq M\sum_{l=1}^I\big(|\partial^1G_l|+|G_l|\big)+M\sum_{m=1}^J|H_j|,\\
&|\underline{L}(H_j)|\leq M\sum_{l=1}^I\big(|\partial^1G_l|+|G_l|\big)+M\sum_{m=1}^J|H_j|,
\end{split}
\end{equation}
for some large constant $M$. We add inequalities \eqref{va10} and \eqref{va10.1} over $i,j$. The key observation is that, in view of \eqref{va11}, the first terms in the right-hand sides of \eqref{va10} and \eqref{va10.1} can be absorbed into the left-hand sides for $\lambda$ sufficiently large. Thus, for any $\lambda$ sufficiently large and $\delta \in(0,1]$,
\begin{equation}\label{va12}
\begin{split}
&\lambda\sum_{i=1}^I\|e^{-\lambda f_{\eps}}\cdot \widetilde{\eta}_{\delta,\eps}G_i\|_{L^2}+\sum_{j=1}^J\|e^{-\lambda f_{\eps}}\cdot \widetilde{\eta}_{\delta,\eps}H_j\|_{L^2}\leq \widetilde{C}\lambda^{-1}\sum_{j=1}^J\|e^{-\lambda f_{\eps}}\cdot H_j|\partial \widetilde{\eta}_{\delta,\eps}|\|_{L^2}\\
&+\widetilde{C}\sum_{i=1}^I\Big[\|e^{-\lambda f_{\eps}}\cdot \D_\al G_i\D^\al \widetilde{\eta}_{\delta,\eps} \|_{L^2}+\|e^{-\lambda f_{\eps}}\cdot G_i( |\square_\g\widetilde{\eta}_{\delta,\eps}|+|\partial\widetilde{\eta}_{\delta,\eps}| )\|_{L^2}\Big].
\end{split}
\end{equation}
We let now $\delta\to 0$ and $\lambda\to\infty$, as in \cite[Section 6]{IoKl}, to conclude that $\mathbf{1}_{B_{\eps^{40}}(x_0)\cap I^{+-}}\, G_i=0$ and $\mathbf{1}_{B_{\eps^{40}}(x_0)\cap I^{+-}}\, H_j=0$. The main ingredient needed for this limiting procedure is the inequality
\begin{equation*}
\inf_{B_{\eps^{40}}(x_0)\cap I_c^{+-}}\,e^{-\lambda f_{\eps}}\geq e^{\lambda/\widetilde{C}}\sup_{\{x\in B_{\eps^{10}}(x_0)\cap I_c^{+-}:N^{x_0}\geq \eps^{20}/2\}}\,e^{-\lambda f_{\eps}},
\end{equation*}
which follows easily from the definition \eqref{mw1}. The lemma follows.
\end{proof}

 \end{document}